\documentclass{article}
\usepackage[utf8]{inputenc}

\usepackage{mathptmx}
\usepackage{times}

\usepackage{amsfonts}
\usepackage{amsmath}
\usepackage{amssymb}
\usepackage{stmaryrd}
\usepackage{color}

\newtheorem{Theorem}{Theorem}
\newtheorem{Lemma}{Lemma}
\newtheorem{Postulation}{Postulation}

\title{Is the Chen-Sbert Divergence a Metric?}
\author{Min Chen$^1$ and Mateu Sbert$^2$\\~\\
$^{1}$ University of Oxford, UK; min.chen@oerc.ox.ac.uk\\
$^{2}$ University of Girona, Spain; mateu@ima.udg.edu}

\date{First version: 1 January, 2021}

\begin{document}

\maketitle

\section{Introduction}
\label{sec:Intro}

Consider any $n$-letter alphabet $\mathbb{Z} = \{z_1, z_2, \ldots, z_n \}$ associated with two probability mass functions, $P = \{ p_1, p_2, \ldots, p_n \}$ and $Q = \{ q_1, q_2, \ldots, q_n \}$. Chen and Sbert proposed a general divergence measure \cite{Chen:2019:arXiv} as follows:

\begin{equation}\label{eq:DCS}
    \mathcal{D}_\text{CS}(P\|Q) = \frac{1}{2}\sum_{i=1}^n (p_i + q_i)
    \log_2 \bigl( |p_i - q_i|^k + 1 \bigr)
\end{equation}
\noindent where $k>0$ is a parameter for moderating the impact of the pairwise-difference between $p_i$ and $q_i$ in relation to other pairwise differences, i.e., $|p_j - q_j|, \forall j \neq i$. Here we focus on the base-2 logarithm in the context of computer science and data science. The transformation to other logarithmic bases is not difficult.

The commonly-used Kullback-Leibler divergence \cite{Kullback:1951:AMS} computes first the informative quantity of individual probabilistic values (in $P$ and $Q$) associated with each letter $z_i \in \mathbb{Z}$. It then computes the pairwise difference between the informative quantities $\log_2 p_i$ and $\log_2 q_i$ for each letter, and finally computes the probabilistic average of such differences $(\log_2 p_i - \log_2 q_i)$, at the informative scale, across all letters $z_i \in \mathbb{Z}$.

Unlike the Kullback-Leibler divergence, $\mathcal{D}_\text{CS}(P\|Q)$ computes first the pairwise difference between $p_i$ and $q_i$ for each letter $z_i \in \mathbb{Z}$, then the informative quantity of the pairwise difference $|p_i - q_i|$ with a monotonic transformation $g(|p_i - q_i|) = \log_2(|p_i - q_i|+1)$, and finally the probabilistic average of such informative quantities across all letters $z_i \in \mathbb{Z}$.

In comparison with the Kullback-Leibler divergence, $\mathcal{D}_\text{CS}(P\|Q)$ is bounded by 0 and 1 (cf. KL is unbounded); does not suffer from any singularity condition (e.g., $q_i = 0$); and is commutative (cf. KL is not commutative). These properties of $\mathcal{D}_\text{CS}(P\|Q)$ are similar to that of the Jensen-Shannon divergence. In an early report \cite{Chen:2019:arXiv}, Chen and Sbert found that some empirical similarities between Jensen-Shannon divergence \cite{Lin:1991:TIT} and $\mathcal{D}_\text{CS}(P\|Q)$ (when $k=2$).

Neither the Kullback-Leibler divergence nor the Jensen-Shannon divergence are a distance metric, but the square root of the Jensen–Shannon divergence is a metric \cite{Endres:2003:TIT,Osterreicher:2003:AISM,Fuglede:2004:PIT}.
This report is concerned with the question whether $\mathcal{D}_\text{CS}(P\|Q)$ is known to be a distance metric. Clearly the measure satisfies most conditions of a distance metric, including:

\begin{itemize}
    \item \emph{identity of indiscernibles}:
    $\mathcal{D}_\text{CS}(P\|Q) = 0 \iff P = Q$;
    \item \emph{symmetry} or \emph{commutativity}:
    $\mathcal{D}_\text{CS}(P\|Q) = \mathcal{D}_\text{CS}(Q\|P)$;
    \item \emph{non-negativity} or \emph{separation}:
    $\mathcal{D}_\text{CS}(P\|Q) \geq 0$.
\end{itemize}

However, it is not clear whether it meets the \emph{triangle inequality} condition (also referred to as the \emph{subadditivity} condition). In other words, if the alphabet $\mathbb{Z}$ is associated with three arbitrary probability mass functions $P$, $Q$ and $R$, do we have the following:

\begin{itemize}
    \item \emph{triangle inequality}:
    $\mathcal{D}_\text{CS}(P\|R) \leq \mathcal{D}_\text{CS}(P\|Q) + \mathcal{D}_\text{CS}(Q\|R)$?
\end{itemize}

\section{Random Testing and Postulations}
\label{sec:Random}

When setting $k=1$ and $k=0.5$ in Eq.\,\ref{eq:DCS}, testing the triangle inequality with randomly generated $P$, $Q$, and $R$ indicates such a possibility.

When setting $k=2$ in Eq.\,\ref{eq:DCS}, testing the triangle inequality with randomly generated $P$, $Q$, and $R$ finds failures of the triangle inequality.
For example, in the case of $n=3$, $P=\{ 0.238, 0.013, 0.749\}$, $Q=\{ 0.253, 0.223, 0.524\}$, and  $R=\{ 0.511, 0.418, 0.071\}$, we have:
\begin{align*}
    &\mathcal{D}_\text{CS}(P \| Q) = \mathcal{D}_\text{CS}(Q \| P) = 0.052\\
    &\mathcal{D}_\text{CS}(Q \| R) = \mathcal{D}_\text{CS}(R \| Q) = 0.133\\
    &\mathcal{D}_\text{CS}(R \| P) = \mathcal{D}_\text{CS}(P \| R) = 0.310\\
    \Longrightarrow \quad
    &\mathcal{D}_\text{CS}(P\|Q) + \mathcal{D}_\text{CS}(Q\|R) - \mathcal{D}_\text{CS}(P\|R) = -0.124 < 0
\end{align*}

Another example of failure is when $n=4$, $P=\{0.143, 0.282, 0.326, 0.248\}$, $Q=\{0.260, 0.172, 0.300, 0.268\}$, and  $R=\{0.040, 0.658, 0.215, 0.088\}$. The calculation shows:
\begin{align*}
    &\mathcal{D}_\text{CS}(P \| Q) = \mathcal{D}_\text{CS}(Q \| P) = 0.008\\
    &\mathcal{D}_\text{CS}(Q \| R) = \mathcal{D}_\text{CS}(R \| Q) = 0.148\\
    &\mathcal{D}_\text{CS}(R \| P) = \mathcal{D}_\text{CS}(P \| R) = 0.102\\
    \Longrightarrow \quad
    &\mathcal{D}_\text{CS}(Q\|P) + \mathcal{D}_\text{CS}(P\|R) - \mathcal{D}_\text{CS}(Q\|R) = -0.038 < 0
\end{align*}
Similar failures have been found when $k=1.5$ and $k=50$. One expects to find many cases of failures with other settings of $k > 1$.\\

Based on the random tests, we propose the following two postulations: 

\begin{Postulation}\label{P:K1}
When $0 < k \leq 1$,
$\mathcal{D}_\text{CS}(P\|R) \leq \mathcal{D}_\text{CS}(P\|Q) + \mathcal{D}_\text{CS}(Q\|R)$.
\end{Postulation}

\begin{Postulation}\label{P:K2}
When $k > 1$,
$\sqrt[k]{\mathcal{D}_\text{CS}(P\|R)} \leq \sqrt[k]{\mathcal{D}_\text{CS}(P\|Q)} + \sqrt[k]{\mathcal{D}_\text{CS}(Q\|R)}$.
\end{Postulation}

\section{Special Case A: $\mathcal{D}_\text{CS}$ ($k=1$) for 2-Letter Alphabets}
\label{sec:2LetterK1}
Consider a 2-letter alphabet $\mathbb{Z}=\{z_1, z_2\}$ and three probability mass functions $P=\{p, 1-p\}$, $Q=\{q, 1-q\}$, and $R=\{r, 1-r\}$.
When $k=1$, the Chen-Sbert measure is simplified as:
\begin{equation}\label{eq:DCS-bi}
\begin{split}
    \mathcal{D}_\text{CS-bi}(P\|Q) &= \mathcal{D}_\text{CS-bi}(Q\|P)\\ 
    &= \frac{1}{2} \biggl(
    (p+q) \log_2 \bigl( |p-q|+1 \bigr) + (2-p-q) \log_2 \bigl( |p-q|+1 \bigr)
    \biggr)\\
    &= \frac{1}{2} \biggl(2 \log_2 \bigl( |p-q|+1\bigr) \biggr) = \log_2 \bigl( |p-q|+1 \bigr)
\end{split}
\end{equation}
Similarly, we have
\begin{align}
    \mathcal{D}_\text{CS-bi}(Q\|R) = \mathcal{D}_\text{CS-bi}(R\|Q) = \log_2 \bigl( |q-r|+1 \bigr)\\
    \mathcal{D}_\text{CS-bi}(R\|P) = \mathcal{D}_\text{CS-bi}(P\|R) = \log_2 \bigl( |p-r|+1 \bigr)
\end{align}
%
%
%

\begin{Lemma}\label{L:K1}
For any real values $0 \leq p, q, r \leq 1$, the following inequality is true:
\begin{equation}\label{eq:TI-bi}
    T(p, q, r) = \log_2 \bigl( |p-q|+1 \bigr) + \log_2 \bigl( |q-r|+1 \bigr) - \log_2 \bigl( |p-r|+1 \bigr) \geq 0
\end{equation}
\end{Lemma}

\noindent\textbf{Proof.} The left side of Eq.\,\ref{eq:TI-bi} can be rewritten as:
\begin{align}
    T(p, q, r) &= \log_2 \frac{(|p-q|+1)(|q-r|+1)}{(|p-r|+1)}\\
    &= \begin{cases} \label{eq:SixCases}
    \log_2 \frac{(a+1)(b+1)}{(a+b+1)} & \text{cases (1) and (6)} \\[2mm]
    \log_2 \frac{(c+d+1)(d+1)}{(c+1)} & \text{cases (2), (3), (4), (5)}
    \end{cases}
\end{align}
\noindent where the six cases are defined as:
\begin{enumerate}
    \item When $p \geq q \geq r$: we set $a = p - q$, $b = q - r$;
    \item When $p \geq r \geq q$: we set $c = p - r$, $d = r - q$;
    \item When $q \geq p \geq r$: we set $c = p - r$, $d = q - p$;
    \item When $q \geq r \geq p$: we set $c = r - p$, $d = q - r$;
    \item When $r \geq p \geq q$: we set $c = r - p$, $d = p - q$;
    \item When $r \geq q \geq p$: we set $a = q - p$, $b = r - q$.
\end{enumerate}

Since $0 \leq a, b, c, d \leq 1$, both parts of Eq.\,\ref{eq:SixCases} are non-negative.
For cases (1) and (6), we have $(a+1)(b+1) = (ab + a + b + 1) \geq (a + b + 1)$.
For cases (2), (3), (4), (5), we have $(c+d+1)(d+1) \geq (c+1)$.
Both fractions inside the logarithmic function in Eq.\,\ref{eq:SixCases} are thus $\geq 1$, and therefore $T(p, q, r) \geq 0$.
According to Eq.\,\ref{eq:DCS-bi} and Eq.\,\ref{eq:TI-bi}, we have:
\[
    \mathcal{D}_\text{CS-bi}(P\|Q) + \mathcal{D}_\text{CS-bi}(Q\|R) - \mathcal{D}_\text{CS-bi}(P\|R) = \frac{1}{2}T(p, q, r) \geq 0
\]
$\mathcal{D}_\text{CS-bi}(P\|Q)$ therefore satisfies the triangle inequality condition.
$\blacksquare$

\begin{Theorem}
    For any 2-letter alphabet, the Chen-Sbert divergence measure (when $k=1$) is a metric.
\end{Theorem}

\noindent\textbf{Proof.} The proof can easily be derived from Lemma \ref{L:K1}, together with the other properties of a distance metric discussed in Section \ref{sec:Intro}.
$\blacksquare$

\section{Special Case B: $\mathcal{D}_\text{CS}$ ($k \leq 1$) for 2-Letter Alphabets}
\label{sec:2LetterK01}
We can extend the above special case to any $0 < k \leq 1$ for any 2-letter alphabet.

\begin{Lemma}\label{L:K01}
For any real values $0 \leq p, q, r \leq 1$ and $0 < k \leq 1$, the following inequality is true:
\begin{equation}\label{eq:TI-bi-k}
    T(p, q, r) = \log_2 \bigl( |p-q|^k+1 \bigr) + \log_2 \bigl( |q-r|^k+1 \bigr) - \log_2 \bigl( |p-r|^k+1 \bigr) \geq 0
\end{equation}
\end{Lemma}

\noindent\textbf{Proof.}
The left side of Eq.\,\ref{eq:TI-bi-k} can be rewritten as:
\begin{align}
    T(p, q, r) &= \log_2 \frac{(|p-q|^k+1)(|q-r|^k+1)}{(|p-r|^k+1)} \label{eq:LogK01}\\
    &= \begin{cases} \label{eq:SixCases01}
    \log_2 \frac{(a^k+1)(b^k+1)}{((a+b)^k+1)} & \text{cases (1) and (6)} \\[2mm]
    \log_2 \frac{((c+d)^k+1)(d^k+1)}{(c^k+1)} & \text{cases (2), (3), (4), (5)}
    \end{cases}
\end{align}
where the six cases are the same as those in Section \ref{sec:2LetterK1}.
Since $0 \leq c, d \leq 1$, it is straightforward to conclude that $( (c+d)^k + 1 )(d^k + 1) \geq (c^k + 1)$ for cases (2), (3), (4), and (5).
Meanwhile, for cases (1) and (6), we consider two terms, $X = a^k b^k + a^k + b^k$ and $Y = (a+b)^k$. If we can show that $X \geq Y$, we will be able to prove the following:
\begin{equation}\label{eq:Reason}
    \frac{X}{Y} \geq 1 \Longrightarrow \frac{X+1}{Y+1} \geq 1 \Longrightarrow
    \frac{a^k b^k + a^k + b^k + 1}{(a+b)^k + 1} = \frac{(a^k + 1)(b^k + 1)}{(a+b)^k + 1} \geq 1    
\end{equation}
We attempt a proof by contradiction. Supposing that $X < Y$, we would have
\begin{equation}\label{eq:Assume}
    \begin{split}
         X < Y &\Longrightarrow a^k b^k + a^k + b^k < (a + b)^k\\
         &\Longrightarrow \bigl( a^k b^k + a^k + b^k \bigr)^{1/k} < (a + b)\\
         &\Longrightarrow \bigl( \sqrt[t]{ab} + \sqrt[t]{a} + \sqrt[t]{b} \bigr)^t < (a + b)
    \end{split}
\end{equation}
\noindent where $t = 1/k > 1$ because $k <1$. Since $0 \leq a, b \leq 1$, we would have:
\begin{equation*}
    (ab + a + b) \leq \bigl( \sqrt[t]{ab} + \sqrt[t]{a} + \sqrt[t]{b} \bigr)^t
\end{equation*}
%

Based on the supposition of $X < Y$ and Eq.\,\ref{eq:Assume}, this would lead to the following conclusion:
\[
    (ab + a + b) \leq \bigl( \sqrt[t]{ab} + \sqrt[t]{a} + \sqrt[t]{b} \bigr)^t < (a + b) \Longrightarrow ab < 0
\]
This contradicts the fact that $0 \leq a, b \leq 1$. Hence $X < Y$ cannot be true.
Because $X \geq Y$ is true, the fraction in the second part of Eq.\,\ref{eq:SixCases01} is also $\geq 1$. For $T(p, q, r)$ in Eq.\,\ref{eq:TI-bi-k}, we can now conclude $T(p, q, r) \geq 0$, and $\mathcal{D}_\text{CS-bi}$ ($0 < k \leq 1$) satisfies the triangle inequality condition.
$\blacksquare$

\begin{Theorem}
    For any 2-letter alphabet, the Chen-Sbert divergence measure (with $0 < k \leq 1$) is a metric.
\end{Theorem}

\noindent\textbf{Proof.} The proof can easily be derived from Lemma \ref{L:K01}, together with the other properties of a distance metric discussed in Section \ref{sec:Intro}.
$\blacksquare$

\section{Special Case C: $\mathcal{D}_\text{CS}$ ($k = 1$) for $n$-Letter Alphabets}
\label{sec:nLetterK1}
 When $k=1$, the might-be triangle inequality of the Chen-Sbert divergence can be expressed as: 
\begin{align*}
    &\mathcal{D}_\text{CS}(P\|Q) + \mathcal{D}_\text{CS}(Q\|R) - \mathcal{D}_\text{CS}(P\|R)
    = \frac{1}{2} \biggl(
        \sum_{i=1}^n (p_i + q_i) \log_2 (|p_i - q_i| + 1)\\
    & \quad + \sum_{i=1}^n (q_i + r_i) \log_2 (|q_i - r_i| + 1) -
        \sum_{i=1}^n (p_i + r_i) \log_2 (|p_i - r_i| + 1) \biggr)\\
    = & \frac{1}{2} \log_2
   \frac{\prod_{i=1}^n \big( |p_i - q_i| +1 \bigr)^{p_i+q_i}
          \big( |q_i - r_i| +1 \bigr)^{q_i+r_i}}%
   {\prod_{i=1}^n \big( |p_i - r_i| +1 \bigr)^{p_i+r_i}} \geq 0
\end{align*}
Proving or falsifying this special case appears to be more difficult.

\subsection{Unsuccessful Pathway: The Individual Term for Each Letter}
This pathway attempts to find a simple proof by examining whether the individual term associated with each letter in the overall summation satisfies the triangle inequality. In other words, we ask a question: 
Is the following inequality always true?
\begin{equation}\label{eq:TermK1}
    \begin{split}
        &(p + q)\log_2(|p-q|+1) + (q + r)\log_2(|q-r|+1) - (p + r)\log_2(|p-r|+1)\\
    = &\log_2 \frac{(|p-q|+1)^{(p+q)} (|q-r|+1)^{(q+r)}}{(|p-r|+1)^{(p+r)}} \geq 0
    \end{split}
\end{equation}
Random tests show that this does not seem to be always true. For example, when $p=0.05, q=0.01, r=0.85$, Eq.\,\ref{eq:TermK1} yields $-0.0016$. In fact, random tests also find instances of negativity when $k < 1$.

However, this is only one term associated with an individual letter (among $n>2$ letters).
The negativity may be cancelled by the terms associated with other letters. For example, for a 2-letter alphabet, when the first letter is associated with $p=0.05, q=0.01, r=0.85$, and the second letter must be associated with $p=0.95, q=0.99, r=0.15$. Eq.\,\ref{eq:TermK1} yields 0.0899 for the second letter. The sum of $-0.0016$ and 0.0899 yields 0.0883, which is still positive.

Hence, the possible negativity for an individual letter is not sufficient for concluding that $\mathcal{D}_\text{CS}$ ($k = 1$) is not a metric for $n$-letter alphabets ($n > 2$).
Nevertheless it at least indicates that if there is a proof, it may not be simple.

\subsection{Unsuccessful Pathway: Combining the Terms of Two Letters}
This pathway attempts to find a proof using induction by examining whether there is a positivity/negativity pattern when combining the terms associated with any two letters.
In other words, let
\[
    T(p, q, r) = (p + q)\log_2(|p-q|+1) + (q + r)\log_2(|q-r|+1) - (p + r)\log_2(|p-r|+1)
\]
and we ask a question: Is the following inequality always true?
\begin{equation}\label{eq:TermK1Add}
    T(p_a + p_b, q_a + q_b, r_a + r_b) \leq T(p_a, q_a, r_a) + T(p_b, q_b, r_b)
\end{equation}
where $0 \leq p_a, p_b, q_a, q_b, r_a, r_b \leq 1$ and $0 \leq (p_a + p_b), (q_a + q_b), (r_a + r_b) \leq 1$.

Random tests show that this is often not true. For example, when $p_a=0.3907, p_b =0.2422, q_a = 0.1134, q_b = 0.0358, r_a = 0.3525, r_b = 0.4558$, Eq.\,\ref{eq:TermK1Add} yields $0.4043$ on the left and $0.2055$ on the right. Hence, this attempt is not successful.

\subsection{Possible Pathway: Replacing the Terms of Three Letters}
\label{sec:3Terms}

Using the same definition of $T(p, q, r)$ as in the previous subsection:
\begin{align*}
    T(p, q, r) &= (p + q)\log_2(|p-q|+1) + (q + r)\log_2(|q-r|+1) - (p + r)\log_2(|p-r|+1)\\
    &= \log_2 \frac{\bigl( |p-q|+1 \bigr)^{p+q} \; \bigl( |q-r|+1 \bigr)^{q+r}}{ \bigl( |p-r|+1 \bigr)^{p+r}}
\end{align*}
this pathway attempts to reduce any three terms $T(p_1, q_1, r_1)$, $T(p_2, q_2, r_2)$, $T(p_3, q_3, r_3)$ associated with three letters to two terms $T(p_x, q_x, r_x)$ and $T(p_y, q_y, r_y)$, such that
\begin{equation}\label{eq:Reduce}
    T(p_1, q_1, r_1) + T(p_2, q_2, r_2) + T(p_3, q_3, r_3) = T(p_x, q_x, r_x) + T(p_y, q_y, r_y)
\end{equation}
where
\begin{equation}\label{eq:Limits}
  \begin{split}
    0 \leq p_1, p_2, p_3, p_x, p_y \leq 1, \quad & 0 \leq p_1 + p_2 + p_3 = p_x + p_y \leq 1\\
    0 \leq q_1, q_2, q_3, q_x, q_y \leq 1, \quad & 0 \leq q_1 + q_2 + q_3 = q_x + q_y \leq 1\\
    0 \leq r_1, r_2, r_3, r_x, r_y \leq 1, \quad & 0 \leq r_1 + r_2 + r_3 = r_x + r_y \leq 1
  \end{split}
\end{equation}



Experiments suggest that this may be achieved by using the following algorithmic steps:
\begin{enumerate}
    \item Initiate three parameters $\alpha = 0$, $\beta = 0$, and $\gamma = 0$.
    \item Initiate $p_x = p_1 + \frac{1}{2}p_3 + \alpha$, \;
    $q_x = q_1 + \frac{1}{2}q_3 + \beta$, \; $r_x = r_1 + \frac{1}{2}r_3 + \gamma$.
    \item Initiate $p_y = p_2 + \frac{1}{2}p_3 - \alpha$, \;
    $q_y = q_2 + \frac{1}{2}q_3 - \beta$, \; $r_y = r_2 + \frac{1}{2}r_3 - \gamma$.
    \item Use an optimisation algorithm to adjust $\alpha$, $\beta$, $\gamma$ and to obtain optimised $(p_x, p_y)$, $(q_x, q_y)$, and $(r_x, r_y)$ such that the requirements in Eqs.\,\ref{eq:Reduce} and \ref{eq:Limits} are met.
\end{enumerate}
%

For example, given:
\[
    (p_1, p_2, p_3) = (0.5, 0.1, 0.2); \; (q_1, q_2, q_3) = (0.1, 0.2, 0.4); \; (r_1, r_2, r_3) = (0.3, 0.3, 0.1)
\]
\[
    \Longrightarrow \quad T(p_1, q_1, r_1) + T(p_2, q_2, r_2) + T(p_3, q_3, r_3) = 0.4967
\]
Following the above algorithmic steps, we can obtain:
\begin{align*}
    p_x = p_1 + \frac{1}{2}p_3 + \alpha = 0.6 + \alpha; \quad &p_y = p_2 + \frac{1}{2}p_3 - \alpha = 0.2 - \alpha\\
    q_x = q_1 + \frac{1}{2}q_3 + \beta = 0.3 + \beta; \quad &q_y = q_2 + \frac{1}{2}q_3 - \beta = 0.4 - \beta\\
    r_x = r_1 + \frac{1}{2}r_3 + \gamma = 0.35 + \gamma; \quad &r_y = r_2 + \frac{1}{2}r_3 - \gamma = 0.35 - \gamma\\
    \Longrightarrow T(p_x, q_x, r_x) + T(p_y, q_y, r_y) &= 0.4967
\end{align*}
We have found the following solution to meet the requirements of Eqs.\,\ref{eq:Reduce} and \ref{eq:Limits}:
\[
    -0.1668334 < \alpha < -0.1668333, \quad \beta = -0.125, \quad \gamma = -0.04
\]
Similarly, we have found solutions for many other instances, e.g.,:
\begin{align*}
    &(p_1, p_2, p_3) = (0.1, 0.2, 0.2); \; (q_1, q_2, q_3) = (0.0, 0.0, 1.0); \; (r_1, r_2, r_3) = (0.2, 0.7, 0.1)\\
    &\qquad \Longrightarrow
    \alpha = -0.14, \quad 0.4161126 < \beta < 0.4161127, \quad \gamma = -0.14\\
    &(p_1, p_2, p_3) = (0.1, 0.9, 0.0); \; (q_1, q_2, q_3) = (0.9, 0.1, 0.0); \; (r_1, r_2, r_3) = (0.2, 0.2, 0.5)\\
    &\qquad \Longrightarrow
    \alpha = -0.05, \quad \beta = 0.05, \quad 0.0442517 < \gamma < 0.0442518
\end{align*}
If an acceptable solution for $\alpha$, $\beta$, and $\gamma$ can always be found, we can use induction to prove that $\mathcal{D}_\text{CS}$ (when $k=1$) is a metric for any $n$-letter alphabet ($n > 0$).

However, we have not made adequate progress beyond that point.

\section{General Case: $\mathcal{D}_\text{CS}$ ($0 < k \leq 1$) for $n$-Letter Alphabets}
\label{sec:nLetterK01}
The strategy for replacing the triangle inequality terms for three letters with equivalent terms of two letters can also be applied to the general case of $\mathcal{D}_\text{CS}$ ($0 < k \leq 1$) for $n$-letter alphabets ($n > 0$).
Our experiments on a number of instances when $k=0.2, 0.5, 0.8$ have successfully found solutions. Nevertheless, a formal mathematical proof will be necessary.

If we can find a proof for special case C ($k=1$) in Section \ref{sec:3Terms}, i.e., there is always a solution of $\alpha$, $\beta$, and $\gamma$ that meets the requirements of Eqs.\,\ref{eq:Reduce} and \ref{eq:Limits}, it is likely that such a proof can be extended to the general case.
In general, it is rather hopeful that Postulation \ref{P:K1} can proved.

\section{Interim Conclusions}
\begin{itemize}
    \item For Postulation \ref{P:K1} in Section \ref{sec:Random}, we have managed to obtain a partial proof that $\mathcal{D}_\text{CS}(P\|Q)$ ($0 < k \leq 1$) is a metric for 2-letter alphabets.
    \item Our attempts to obtain a full proof for the general case of $n$-letter alphabets have only found a possible but unconfirmed pathway.
    \item We have not yet made any solid progress in proving or falsifying Postulation \ref{P:K2} in Section \ref{sec:Random}.
\end{itemize}

We shall appreciate any effort to prove or falsify the two postulations by colleagues in the international scientific communities.

\bibliographystyle{abbrv}
\nocite{*}
\bibliography{metric}

\end{document}